\documentclass[aps,prl,reprint, amsmath, amssymb,superscriptaddress,nofootinbib]{revtex4-1}

\usepackage{bm}
\usepackage[retainorgcmds]{IEEEtrantools}
\usepackage{graphicx}
\usepackage{mathrsfs}
\usepackage{amsmath}
\usepackage{amssymb}
\usepackage{color}
\usepackage{amsfonts}
\usepackage{times,txfonts}
\usepackage{nicefrac}
\usepackage[colorlinks=true,linkcolor=blue,urlcolor=blue,citecolor=blue,pdfusetitle]{hyperref}

\usepackage{soul}

\usepackage[dvipsnames]{xcolor}
\definecolor{mypink}{rgb}{0.858, 0.188, 0.478}
\usepackage[authormarkup=none]{changes}
\definechangesauthor[name={Gabriel}, color=red]{g}

\DeclareMathOperator{\tr}{tr}

\begin{document}

\title{Landauer's principle at zero temperature}
\date{\today}
\author{Andr\'e M. Timpanaro}
\affiliation{Universidade Federal do ABC,  09210-580 Santo Andr\'e, Brazil}
\author{Jader P. Santos}
\affiliation{Instituto de F\'isica da Universidade de S\~ao Paulo,  05314-970 S\~ao Paulo, Brazil.}
\author{Gabriel T. Landi}
\email{gtlandi@if.usp.br}
\affiliation{Instituto de F\'isica da Universidade de S\~ao Paulo,  05314-970 S\~ao Paulo, Brazil.}

\begin{abstract}

Landauer's bound relates  changes in the entropy of a system with the inevitable dissipation of heat to the environment.
The bound, however, becomes trivial in the limit of zero temperature.
Here we show that it is possible to derive a tighter bound which remains non-trivial even as $T\to 0$.
As in the original case, the only assumption we make is that the environment is in a thermal state. Nothing is said about the state of the system or the kind of system-environment interaction. 
Our bound is valid for all temperatures and is always tighter than the original one, tending to it in the limit of high temperatures.

\end{abstract}
\maketitle{}

{\bf \emph{Introduction - }}
Around six decades ago Landauer showed that erasing  information in a memory has a fundamental heat cost \cite{Landauer1961}. 
This can be stated mathematically as a bound comparing the heat $\Delta Q_E$ dissipated to the environment with the change in entropy $\Delta S_S$ of the system, viz.,
\begin{equation}\label{landauer}
\Delta Q_E \geq - T \Delta S_S,
\end{equation}
where $T$ is the temperature of the environment\footnote{Here and henceforth $\Delta$'s always refer to ``final minus initial``, unlike Ref.~\cite{Reeb2014} which uses ``initial minus final''.}. 
This result is of practical  relevance, as it provides  guidelines  on the ultimate dissipative costs of computation.
This is particularly relevant when $\Delta S_S < 0$,  in which case it bounds the minimum heat cost necessary for purifying the state of the system.
Eq.~(\ref{landauer}) is also of fundamental interest, establishing a deep connection between thermodynamics and information: To manipulate information one has to pay a price in dissipation~\cite{Sagawa2009}.

In view of the growing interest in quantum information sciences, the extension of Landauer's principle to the quantum regime has seen a boom of interest in the last decade. 
In particular, it was  shown in \cite{Esposito2010a,Reeb2014} that the bound~(\ref{landauer}) is a direct consequence of the second law of thermodynamics, valid arbitrarily far from equilibrium. 
In this scenario a system $S$ interacts with an environment $E$ by means of a global unitary $U$, generating a map
\begin{equation}
    \rho_{SE}' = U \bigg(\rho_S \otimes \rho_E\bigg) U^\dagger, 
    \label{global_map}
\end{equation}
where $\rho_{S(E)}$ are the initial states of the system and bath respectively. 
Quite important, no specific assumptions are made about the interaction $U$ or the initial state of the system $\rho_S$.
The only assumption is that the bath itself is in a thermal state at a temperature $T$; i.e., $\rho_E \equiv \rho_{E}^\text{th}(T) =  e^{-H_E/T}/Z_E$, where $H_E$ is the environment's Hamiltonian and $Z_E$ the partition function ($k_B = 1$). 
The map~(\ref{global_map}) is therefore extremely general.

\begin{figure}[!t]
    \centering
    \includegraphics[width=0.45\textwidth]{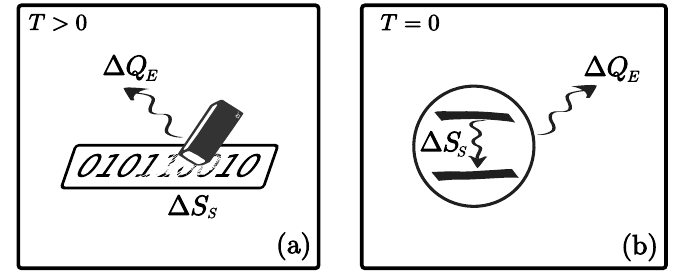}
    \caption{(a) Landauer's principle~(\ref{landauer}) relates the amount of heat an environment must absorb in order to erase information about a system. (b) In the limit of $T\to 0$, however, the bound becomes uninformative. This is unsatisfactory since even simple processes, such as spontaneous emission, fall under this category.}
    \label{fig:drawing}
\end{figure}

The second law associated with~(\ref{global_map}) can be formulated solely in terms of information-theoretic quantities \cite{Esposito2010a}, by defining the entropy production as
\begin{equation}
    \Sigma := I'(S\! : \! E) + S(\rho_E' || \rho_E) \geq 0.
    \label{Sigma}
\end{equation}
The first term is the mutual information developed between $S$ and $E$ due to the interaction, $I'(S\! : \! E) = S(\rho_S') + S(\rho_E') - S(\rho_{SE}')$, where $S(\rho) = - \tr(\rho \ln \rho)$ is the von Neumann entropy. The second, $S(\rho_E'|| \rho_E) = \tr (\rho_E' \ln \rho_E' - \rho_E' \ln \rho_E)$, is the quantum relative entropy between the final non-equilibrium state of the bath, $\rho_E' = \tr_S \rho_{SE}'$ and the initial thermal state. 
Since the map~(\ref{global_map}) is unitary, the mutual information simplifies to 
\begin{equation}
    I'(S\! : \! E) = \Delta S_S + \Delta S_E \geq 0,
    \label{MI}
\end{equation}
Moreover, since $\rho_E = \rho_E^\text{th}(T)$, it follows that
\[
\Delta S_E + S(\rho_E' || \rho_E^\text{th}(T)) = \beta \Delta Q_E := \beta \tr\big\{ H_E(\rho_E' - \rho_E^\text{th}(T))\big\}.
\]
Plugging these results in Eq.~(\ref{Sigma}) immediately yields 
\begin{equation}
    \Sigma = \Delta S_S + \beta \Delta Q_E \geq 0,
\end{equation}
from which the bound~(\ref{landauer}) follows.
Landauer's bound is thus a direct consequence of the second law $\Sigma \geq 0$. 

The above derivation is simple and illuminating.
It also highlights what we believe are the two essential ingredients of the bound~(\ref{landauer}). 
First, it is written solely in terms of $\Delta S_S$ and $\Delta Q_E$. 
The former is the information-theoretic quantity of interest while the latter is an easily accessible quantity of the bath. 
Second, the bound does not require any information about $\rho_S$ or $U$. The only assumption is that the environment is thermal. 
This is quite relevant since, if one knows all there is to know about $S+E$,  having a bound is not really necessary; one can simply calculate the exact amount of dissipated heat. 
Landauer's bound is useful \emph{precisely} because it is universal and requires minimal information.

In recent years, several generalizations of~(\ref{landauer}) have  been put forth. 
In Ref.~\cite{Goold2014b} the authors used the exchange fluctuation theorem to make the bound tighter, although it required knowledge of the system-environment  unitary $U$. 
Similarly, in Ref.~\cite{Guarnieri2017}  an entire family  of bounds was derived using full-counting statistics. 
The extension to collisional models was put forth in
Refs.~\cite{Lorenzo2015,Strasberg2016} and the role of quantum coherence was studied in~\cite{Campbell2017a}.
Experimental demonstrations of Landauer's principle in the microscopic domain were recently given, in a nuclear magnetic resonance setup~\cite{Peterson2016}, a molecular nanomagnet~\cite{Gaudenzi2018}, a Brownian particle~\cite{Toyabe2010} and a trapped ion systems~\cite{Yan2018}.
Generalizations to account for initial system-environment correlations have also been put forth~\cite{Partovi2008,Jennings2010,Jevtic2015a,Micadei2019} and experimentally verified~\cite{Micadei2017}.


In the limit $T\to 0$, however, the bound~(\ref{landauer}) becomes trivial; it simply states that $\Delta Q_E \geq 0$, irrespective of $\Delta S_S$ (this is also true for all generalizations reported above).
The reason why it is trivial is because, when $T = 0$, the bath will be in the ground-state, so that any physical process must always satisfy $\Delta Q_E \geq 0$. 
The bound does remain useful as a way of emphasizing that there can be processes which occur with zero heat cost (i.e., the bound is still saturable). 
But other than that, it does not provide any information, which is clearly unsatisfactory~Ref~\cite{Uzdin2018a,Uzdin2019}.  
Take, as an example, the process of spontaneous emission of an atom into the radiation field (Fig.~\ref{fig:drawing}). 
Any change in entropy of the system will always be accompanied by a flow of heat (represented by the energy carried by a photon). 
It would therefore be interesting to obtain bounds which capture these fine grained effects. 

In this letter we show that it is possible to derive a modified bound which yields non-trivial information even at zero temperature. 
Our bound is summarized by Eq.~(\ref{landauer_new}) below. 
It is always tighter than~(\ref{landauer}) and valid for all temperatures.
And tends to~(\ref{landauer}) in the limit of high temperatures.
But, most crucially, it remains non-trivial as $T\to 0$. 
It also requires only one additional piece of information, namely the environment's specific heat as a function of temperature. 
No information is required about the system-environment interaction or the initial state of the system. 

Written in the form~(\ref{landauer_new}), our new bound is somewhat abstract. But it can be made explicit for specific environments. 
Several examples are discussed. One, which is particularly illuminating,  is that of emission onto a one-dimensional waveguide. 
As we demonstrate below, in this case we find
\begin{equation}\label{waveguide_bound_intro}
    \Delta Q_E \geq - T \Delta S_S + \frac{3 \hbar c}{\pi L} \Delta S_S^2,
\end{equation}
which holds for an arbitrary system coupled in an arbitrary way to the waveguide.
Here $c$ and $L$ are the speed of light and the length of the waveguide respectively. 
The appearance of the second term makes this bound always stricter than~(\ref{landauer}). Moreover, it remains informative even when $T\to 0$. 
In particular, it shows that it is impossible to change $\Delta S_S$ without an ensuing heat exchange $\Delta Q_E$. 


{\bf \emph{Derivation of the modified bound - }} Our starting point is the general quantum map~(\ref{global_map}). 
The two terms in the second law~(\ref{Sigma}) are individually non-negative.
The reason why~(\ref{landauer}) becomes uninformative when $T\to 0$ is because  the last term in~(\ref{Sigma}) diverges when the support of $\rho_E$ is not contained in that of $\rho_E'$, which happens because $\rho_E$ tends to a pure state. 
The mutual information~(\ref{MI}), on the other hand, remains finite. 
The key insight in our scheme is to use only the mutual information to derive the bound. 

The initial state of the environment is thermal and thus characterized by an equilibrium entropy $S_E(T) = S(\rho_E^\text{th}(T))$ and internal energy $E_E(T)  = \tr\big\{H_E \rho_E^\text{th}(T)\big\}$.
The final bath $\rho_E'$, on the other hand, is generally far from equilibrium.
Let us then introduce a reference thermal state $\rho_E^\text{th}(T')$, at a temperature $T'$ chosen such that it has the same internal energy as the final state $\rho_E'$; i.e.,  
\[
\tr\big\{H_E \rho_E' \big\} = \tr\big\{ H_E \rho_E^\text{th}(T')\big\} = E_E(T').
\]
Since internal energy is always in one-to-one with temperature, the value of $T'$ is unique (although possibly negative). 

According to the MaxEnt principle, out of all states of $E$ having energy $E_E(T')$, the thermal state $\rho_E^\text{th}(T')$ is the one with the highest possible von Neumann entropy. 
Whence,
\begin{equation}\label{MaxEnt}
S_E(T') = S(\rho_E^\text{th}(T'))  \geq S(\rho_E'). 
\end{equation}
Plugging this in Eq.~(\ref{MI}) yields the bound
\begin{equation}\label{intermediate_bound_thermal}
    \Delta S_S + \Delta S_E^\text{th} \geq \Delta S_S + \Delta S_E \geq 0,
\end{equation}
where $\Delta S_E^\text{th} = S_E(T') - S_E(T)$ is now a difference between equilibrium entropies. 

As the final step, we consider the quantities $\Delta Q_E = E_E(T') - E_E(T)$ and $\Delta S_E^\text{th}$ and interpret them as  functions of $\beta' = 1/T'$ (we use $\beta'$ instead of $T'$ merely for convenience):
\begin{equation}\label{functions_Q_and_S}
    \Delta Q_E := \mathcal{Q}(\beta') \quad\text{ and }\quad \Delta S_E^\text{th} := \mathcal{S}(\beta'). 
\end{equation}
The function $\mathcal{Q}(\beta')$ is always monotonically decreasing in $\beta'$ and thus has a unique inverse  $\beta' = \mathcal{Q}^{-1} (\Delta Q_E)$. Hence, we may write 
\begin{equation}
    \Delta S_E^\text{th} = \mathcal{S}(\beta') = \mathcal{S}(\mathcal{Q}^{-1}(\Delta Q_E)).
\end{equation}
Plugging this in Eq.~(\ref{intermediate_bound_thermal}) yields
\begin{equation}\label{landauer_new_0}
     \mathcal{S}(\mathcal{Q}^{-1}(\Delta Q_E))\geq -\Delta S_S.
\end{equation}
This bound is already in the desired form~(\ref{landauer}),  relating $\Delta Q_E$ to $\Delta S_S$. 

Finally, we can cast the result precisely in the same way as Eq.~(\ref{landauer}) by introducing the inverse $\mathcal{S}^{-1}$. 
If the environment does not admit negative temperatures, as is the case when it is infinite dimensional, this inverse is unique.
Conversely, if it admits negative temperatures, the inverse will be double valued, with one solution for $\beta'>0$ and another for $\beta'<0$. 
Notwithstanding, we show in the Supplemental Material~\cite{SupMat} that the lower bound for $\Delta Q_E$ is obtained by taking the solution with $\beta'>0$, which is what we shall henceforth refer to as $\mathcal{S}^{-1}$.
As a result, Eq.~(\ref{landauer_new_0}) can be finally rewritten as
\begin{equation}
    \Delta Q_E \geq \mathcal{Q}\big( \mathcal{S}^{-1}(-\Delta S_S)\big).
    \label{landauer_new}
\end{equation}
This is the main result of this paper. 
It provides a bound on the heat $\Delta Q_E$ absorbed by the bath when the entropy of the system  changes by $\Delta S_S$.
It is identical in spirit to Eq.~(\ref{landauer}): it requires no information on the initial state $\rho_S$ of the system nor on the system-environment unitary $U$. 
Note also that even though $\rho_S$ and $U$ in principle affect $T'$, nowhere do we actually need to know $T'$, which is used merely as an auxiliary variable. 

The difference in comparison to~(\ref{landauer}) is that  the two quantities $\Delta Q_E$ and $\Delta S_S$ are connected here through a less trivial function $\mathcal{Q}(\mathcal{S}^{-1}(\bullet))$, whereas in~(\ref{landauer}) they are connected simply by $-T \bullet$. 
This new function, however, involves only thermal equilibrium quantities of the bath (even though the process is arbitrarily far from equilibrium).
In fact, in terms of the bath's equilibrium heat capacity $C_E(T)$, Eq.~(\ref{functions_Q_and_S}) can be written as~\cite{Landau1958} 
\begin{equation}\label{heat_capacity}
    \mathcal{Q}(T') = \int\limits_T^{T'} C_E(\tau) d\tau, 
    \qquad
    \mathcal{S}(T') = \int\limits_{T}^{T'} \frac{C_E(\tau)}{\tau} d\tau.
\end{equation}
Thus, we see that our bound requires \emph{only} a single function, $C_E(T)$.
This is of course more than the original bound~(\ref{landauer}), which requires only a single number ($T$). 
But still, knowing (or at least having an estimate) of the environment's heat capacity is not too complicated (see below for examples).


{\bf \emph{Comparison with the original bound - }}The bound~(\ref{landauer_new}) is always tighter than~(\ref{landauer}):
\begin{equation}\label{tightness}
    \mathcal{Q}\big(\mathcal{S}^{-1}(-\Delta S_S)\big) \geq - T \Delta S_S.
\end{equation}
Quite elegantly, this can be shown to be a consequence solely of equilibrium thermodynamics. 
Since all functions involved are strictly monotonic, Eq.~(\ref{tightness}) is tantamount to 
\begin{equation}
    \mathcal{Q}(\beta') \geq T \mathcal{S}(\beta').
\end{equation}
Using Eq.~(\ref{functions_Q_and_S}) this can in turn be written as 
\begin{equation}\label{free_energy_bound}
    F_E(\rho_E^\text{th}(T')) \geq F_E(T),
\end{equation}
where $F_E(\rho) = \tr\big\{H_E \rho\big\} - T S(\rho)$
is the non-equilibrium free energy of the environment defined with $T$ (and not $T'$) as a reference temperature and $F_E(T) = F_E(\rho_E^\text{th}(T))$ is the corresponding equilibrium free energy. Eq.~(\ref{free_energy_bound}) is a fundamental property of the free energy~\cite{Fermi1956} (equivalent to the MaxEnt principle): out of all states of $E$, the thermal state $\rho_E^\text{th}(T)$ at a temperature $T$ is the one which minimizes $F_E(\rho)$. 
This can be readily proven by writing
$F_E(\rho) = F_E(T) + T S(\rho || \rho_E^\text{th}(T))$ and using the fact that the relative entropy is always non-negative. 
This therefore proves~(\ref{free_energy_bound}) and consequently~(\ref{tightness}). 

{\bf \emph{Application: Rabi model - }} We now illustrate the applicability of our main result~(\ref{landauer_new}), by considering the simple  example of  spontaneous emission of a two-level atom onto a single-mode cavity, as described by the Rabi model
\begin{equation}\label{Rabi}
    H = \hbar\omega a^\dagger a + \frac{ \hbar\Omega}{2} \sigma_z  +  \hbar g (a+a^\dagger)\sigma_x
\end{equation}
where $a$ is the cavity mode and $\sigma_i$ are Pauli matrices. 
Here the atom plays the role of the system, whereas the cavity plays the role of the environment. 
We assume the initial state of the atom is a simple thermal state with excitation probability $p$, whereas the cavity is in a thermal state at temperature $T$. 

The functions $E_E(T)$ and $S_E(T)$ used to construct $\mathcal{Q}(T')$ and $\mathcal{S}(T')$ in Eq.~(\ref{functions_Q_and_S}) read, in this case, $E_E(T) = \omega \bar{n}(T)$ and $S_E(T) =  (\bar{n}(T)+1) \ln (\bar{n}(T) + 1)- \bar{n}(T) \ln \bar{n}(T)$. 
The function inverse of $\mathcal{S}(T')$ has no analytic form. Notwithstanding, it can be trivially found numerically.

\begin{figure}
    \centering
    \includegraphics[width=0.45\textwidth]{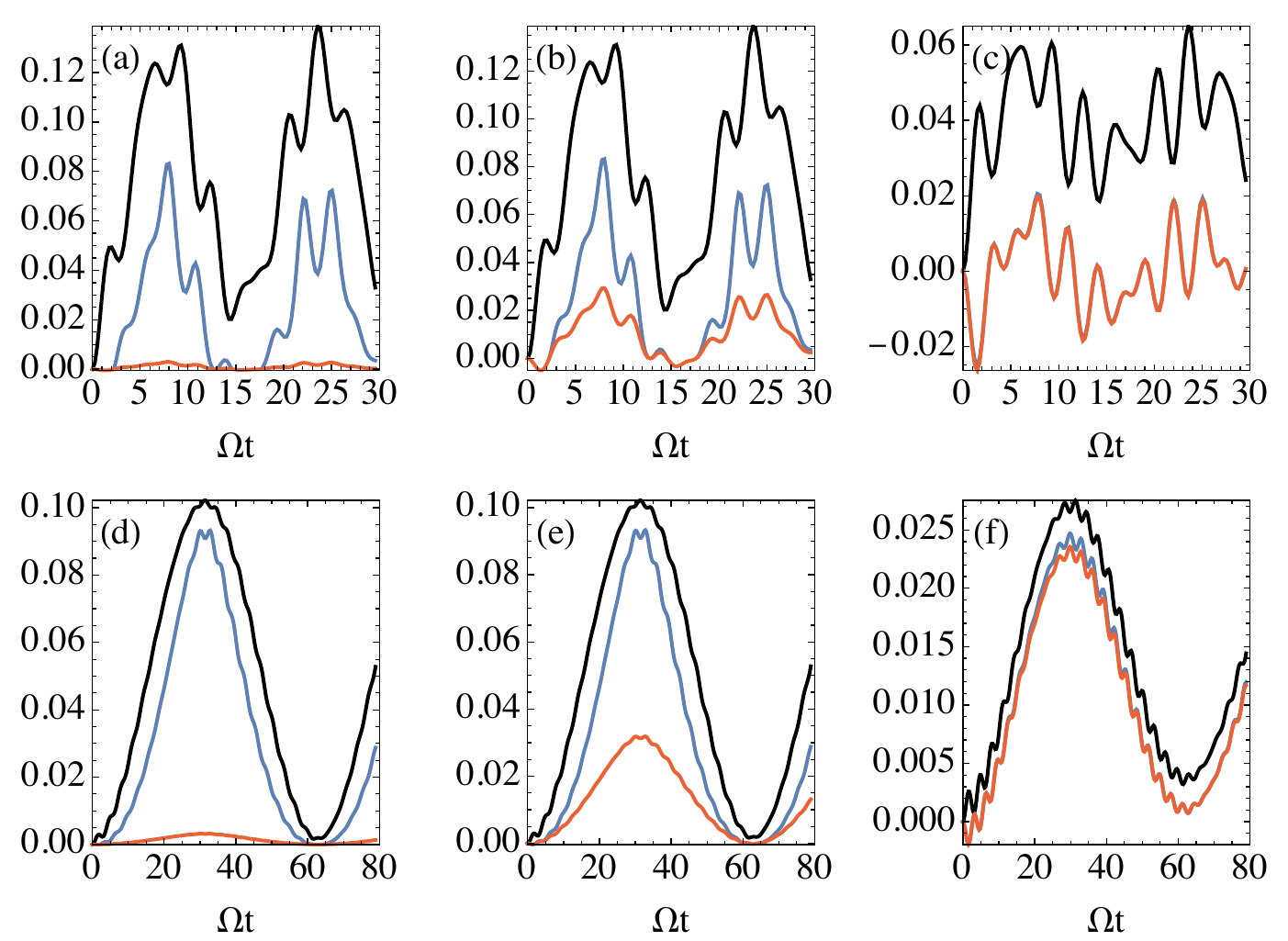}
    \caption{
    Benchmark of the modified bound~(\ref{landauer_new}) for the Rabi model [Eq.~(\ref{Rabi})].
    The plots compare the heat absorbed $\Delta Q_E$ by the cavity mode (top black curves) with the  bound~(\ref{landauer_new}) (middle blue curves) and the original Landauer bound~(\ref{landauer})  (bottom red curves).
    Top row: $g = 0.2$; bottom row: $g = 0.05$. 
    (a,d)  $T = 0.01$; (b,e) $T = 0.1$; (c,f) $T = 0.4$.     
    The qubit was prepared also in a thermal state with excitation probability $p = 0.1$. Other parameters were $\omega = \Omega = 1$. 
    }
    \label{fig:Rabi}
\end{figure}

A comparison of the heat $\Delta Q_E$ absorbed by the cavity mode  and the two bounds~(\ref{landauer}) and (\ref{landauer_new}) is shown in Fig.~\ref{fig:Rabi}. 
The different images are for increasing values of $T$ (from left to right) with large and small  $g$ in the top and bottom rows, respectively.
We call attention to a comparison between the curves for $T = 0.01$ (Fig.~\ref{fig:Rabi}(a,d)) and $T = 0.1$ (Fig.~\ref{fig:Rabi}(b,e)). 
In this parameter range the heat exchange $\Delta Q_E$ is rather insensitive to $T$ (the black curves in (a) and (b) are practically identical; and similarly for (d) and (e)). 
Notwithstanding, the original bound~(\ref{landauer}) becomes less and less tight as the temperature is decreased;
for $T = 0.01$ it is already practically uninformative.
This illustrates well what we believe is the main motivation behind our work: the emission process itself (the black curves in Fig.~\ref{fig:Rabi}) is practically unaffected as the temperature is reduced, whereas the bound~(\ref{landauer}) become increasingly  worse. 
Eq.~(\ref{landauer_new}), on the other hand, follows to a great extent the features of $\Delta Q_E$, being also insensitive to $T$. It  is also considerably tighter than~(\ref{landauer}), specially at low temperatures.
Conversely, for high temperatures the bounds asymptotically coincide (Fig.~\ref{fig:Rabi}(c,f)).  

{\bf \emph{Emission onto a one-dimensional waveguide - }} We now present the derivation of Eq.~(\ref{waveguide_bound_intro}), for the interaction of a system with a one-dimensional waveguide.  
The only assumption we make is that the  waveguide is modeled by a set of bosonic modes $b_k$
with a dispersion relation of the form $\omega_k \simeq c k$, where $c$ is the speed of light. 

The internal energy $E_E(T)$ and the equilibrium entropy $S_E(T)$ can be found analytically by transforming the sum to an integral.
Introducing the Bose-Einstein thermal distribution $\bar{n}_k = (e^{\beta \hbar \omega_k} -1)$, one finds 
\begin{IEEEeqnarray*}{rCl}
E_E(T) &=& \sum\limits_k \hbar \omega_k \bar{n}_k = \frac{\pi L}{12 \hbar c}T^2, \\[0.2cm]
S_E(T) &=& \sum\limits_k \bigg\{ (\bar{n}_k + 1) \ln (\bar{n}_k + 1) - \bar{n}_k \ln \bar{n}_k \bigg\}= \frac{\pi L}{6\hbar c} T,
\end{IEEEeqnarray*}
where $L$ is the waveguide's length.
The functions in Eq.~(\ref{functions_Q_and_S}) are written more simply in this case as a function of $T'$ instead of $\beta'$. 
We then find $\mathcal{S}(T') = \frac{\pi L}{6\hbar c} (T'-T)$, whose inverse is  $T' = T + \frac{6\hbar c}{\pi L} \mathcal{S}$. 
Similarly, $\mathcal{Q}(T') = \frac{\pi L}{12 \hbar c} (T'^2 - T^2)$, so that 
\[
\mathcal{Q}(T') = T \mathcal{S} + \frac{3 \hbar c}{\pi L} \mathcal{S}^2.
\]
Finally, returning to Eq.~(\ref{landauer_new}) and substituting $\mathcal{S} \to - \Delta S_S$, we obtain Eq.~(\ref{waveguide_bound_intro}). 

{\bf \emph{Heat capacity examples - }}
The bound~(\ref{landauer_new}) only requires knowledge of the environment's heat capacity $C_E(T)$. 
With reasonable guesses for $C_E(T)$ one may therefore produce estimates for a variety of systems. 
Phonons, for instance, usually have a heat capacity scaling as $C_E(T) = a T^3$ at low temperatures~\cite{Ashcroft1976}, where $a$ is a constant.
Eq.~(\ref{heat_capacity}) then yields $\mathcal{Q}(T') = \frac{a}{4} (T'^4-T^4)$ and $\mathcal{S}(T') = \frac{a}{3} (T'^3 - T^3)$. 
Plugging this in Eq.~(\ref{landauer_new}) yields a bound depending only on $T$, $\Delta S_S$ and $a$. 
In the limit $T=0$ this bound simplifies to: 
\begin{equation}\label{phonon}
    \Delta Q_E \geq \frac{3^{4/3}}{4} \frac{(-\Delta S_S)^{4/3}}{a^{1/3}}. 
\end{equation}
The minimum heat cost therefore scales as $\Delta S_S^{4/3}$, instead of $\Delta S_S^2$ as in Eq.~(\ref{waveguide_bound_intro}). 

Another interesting example is that of gapped environments. That is, environments for which there is a gap between the ground-state and the first excited state. 
The prime example are BCS superconductors~\cite{Bardeen1957}. 
For such systems, the heat capacity at low temperatures has the form  $C_E(T) = b e^{-\delta/T}$, where $\delta$ is the energy gap and $b$ is a constant. 
In this case, as shown in the Supplemental Material~\cite{SupMat}, the bound~(\ref{landauer_new}) at $T = 0$,  assuming $b \gg -\Delta S_S$, becomes
\begin{equation}\label{gapped}
    \Delta Q_E \geq \delta \frac{(-\Delta S_S)}{\ln (-b/\Delta S_S)}.
\end{equation}
The bound therefore retains a roughly linear dependence, with a logarithmic correction. 

{\bf \emph{Dependence on the environment's size - }}
As in the original  formulation~\cite{Esposito2010a,Reeb2014}, our framework is not restricted to the traditional idea that environments must be infinitely large. 
For us, the environment can have any size. 
The only assumption is that it is thermal. 
Indeed, our framework is better viewed as describing the  interaction between two systems $S$ and $E$ of arbitrary size, one of which ($E$) is in a thermal state. 
This is a significant advantage, since current research in quantum physics  often deals with finite-size environments. 
Non-Markovianity, for instance, relies heavily on it. 
Similarly, a Bose-Einstein condensate  acts as a bath  for an impurity. And this bath is definitely finite. 
Experiments involving optical cavities (c.f. Fig.~\ref{fig:Rabi}) also offer another good illustration.

Notwithstanding, it is interesting to analyze what happens in the limit where the environment is macroscopically large. 
Unlike Eq.~(\ref{landauer}), our bound  depends on the environment's size.
This is evident in Eq.~(\ref{waveguide_bound_intro}), but is also present in Eqs.~(\ref{phonon}) and (\ref{gapped}) since extensivity implies $a,b \sim V_E$, the volume of the environment. 
Thus, when the environment is large, our bound may also require 
that $\Delta S_S$ is comparably large in order to yield a non-negligible correction. 
Precisely how large, however, can only be determined on a case-by-case basis. 
In the case of waveguides, Eq.~(\ref{waveguide_bound_intro}),  one must have $\Delta S_S \sim L^{1/2}$, where $L$ is the size of the waveguide. 
Conversely, for a phonon bath, Eq.~(\ref{phonon}), one requires only $\Delta S_S\sim V_E^{1/4}$, which is much weaker.  

The gapped system example in Eq.~(\ref{gapped}) is the most interesting. 
It requires only $\Delta S_S \sim \ln V_E$.
Such a logarithmic dependence therefore makes this effect visible even for quite small $\Delta S_S$. 
Gapped systems such as superconductors, should thus have a significant heat cost for erasing information in the low temperature regime.

{\bf \emph{Conclusions - }} Landauer's bound is useful because it requires minimal information. However, it becomes uninformative when $T\to 0$. 
In this letter we have derived a new bound, identical in spirit.
The bound is valid for all temperatures and is always tighter.
For high temperatures it coincides with the original one. But most crucially, it remains useful even at $T=0$. 

Our derivation was based on two inequalities: the positivity of the mutual information~(\ref{MI}) and the MaxEnt principle~(\ref{MaxEnt}). 
The former is saturated asymptotically when the system-environment correlations are vanishingly small and the latter when the final state of the environment after interacting with the system is still approximately thermal. 
For macroscopically large and highly chaotic baths, both conditions tend to be met. 
However, many of the environments currently being used in quantum-coherent experiments, such as optical cavities and waveguides, are not of this form. 
This highlights the relevance and timeliness of our results, which provides a route to extend Landauer's principle beyond the standard thermal paradigms.

\emph{Acknowledgements - }
The authors  acknowledge Raam Uzdin for fruitful discussion, as well as for coining the term ``zero temperature catastrophe''. 
The authors acknowledge fruitful discussions with M. Paternostro, F. Brito, J. Goold, L. H. Mandetta, G. Guarnieri and D. Salazar. 
We are particularly grateful to Kamil Korzekwa for pointing out the subtleties involved when the environment is finite dimensional. 
G. T. L. acknowledges the financial support of the S\~ao Paulo Funding Agency FAPESP (Grants No. 2017/50304-7, 2017/07973-5 and 2018/12813-0) and the Brazilian funding agency CNPq (Grant No. INCT-IQ 246569/2014-0).
J. P. S. acknowledges the financial support from the CAPES PNPD program.

\bibliography{library}

\pagebreak
\widetext

\newpage 
\begin{center}
\vskip0.5cm
{\Large Supplemental Material}
\end{center}
\vskip0.4cm

\setcounter{equation}{0}
\setcounter{figure}{0}
\setcounter{table}{0}
\setcounter{page}{1}
\renewcommand{\theequation}{S\arabic{equation}}
\renewcommand{\thefigure}{S\arabic{figure}}

In this supplemental material we provide additional details on how to go from Eq.~(\ref{landauer_new_0}) to Eq.~(\ref{landauer_new}) when the environment allows for negative temperatures (as is the case, e.g., when it is finite dimensional). 
We also provide details on the calculation of Eq.~(\ref{gapped}). 

\section*{Details on how to go from Eq.~(\ref{landauer_new_0}) to Eq.~(\ref{landauer_new})}

The function $\mathcal{Q}(\beta')$ is always a monotonically decreasing function of $\beta'$ and thus has a unique inverse.
But the function $\mathcal{S}(\beta')$ will only be monotonically decreasing when the environment does not allow for negative temperature. 
If it does, there will be two branches, one with $\beta'>0$ and another with $\beta'<0$. 
Our goal here is to show that as far as our modified bound is concerned, one should take the inverse with $\beta'>0$.

\begin{figure}[!h]
    \centering
    \includegraphics[width=0.8\textwidth]{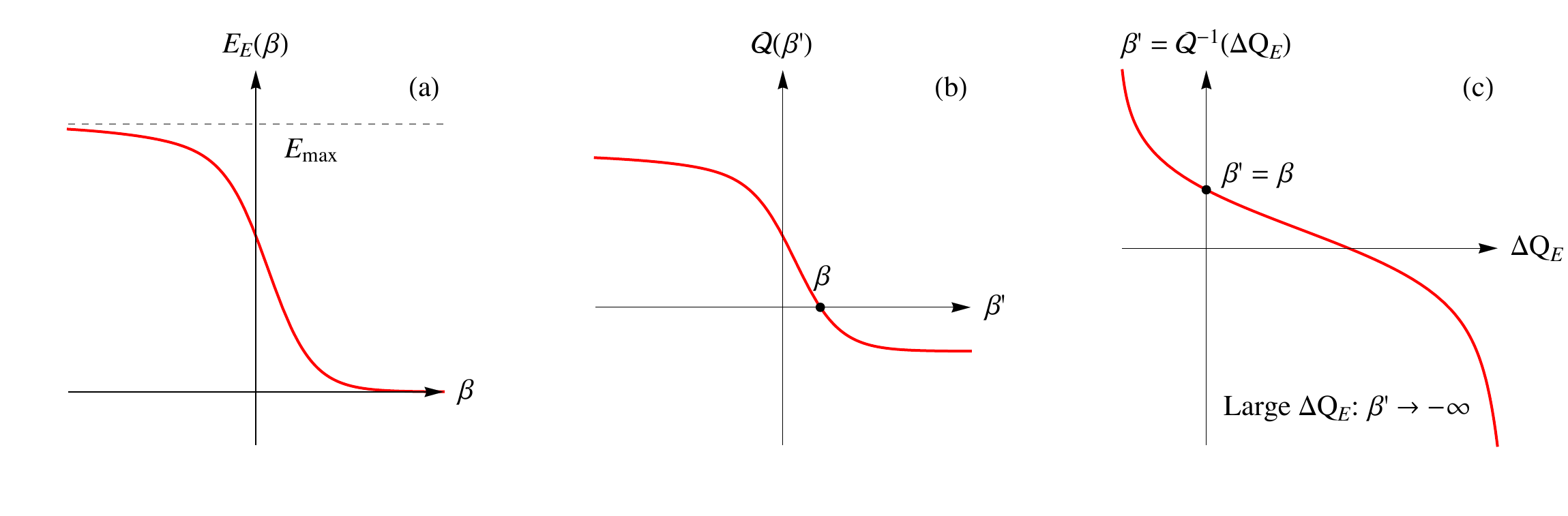}
    \caption{\label{fig:SM_energies}Typical behavior of $E_E(\beta)$, $\mathcal{Q}(\beta')$ and $\mathcal{Q}^{-1}(\Delta Q_E)$ for systems that admit negative temperatures. }
\end{figure}

We illustrate this graphically as follows.
The internal energy when the environment admits negative temperatures has the general form shown in Fig.~\ref{fig:SM_energies}(a). 
It is maximal when $\beta \to -\infty$ and minimal when $\beta \to \infty$. 
The corresponding function $\mathcal{Q}(\beta') = E_E(\beta') - E_E(\beta)$ is then simply a shifted version of the same function (Fig.~\ref{fig:SM_energies}(b)). 
Moreover, this shift is such that $\mathcal{Q}(\beta') = 0$ when $\beta' = \beta$, as shown in the figure. 
Since $\mathcal{Q}(\beta')$ is  monotonically decreasing, it has a unique inverse $\Delta Q_E = \mathcal{Q}^{-1}(\beta')$ (which is also monotonically decreasing), as shown in Fig.~\ref{fig:SM_energies}(c). 
In this image we emphasize that large $\Delta Q_E$ implies $\beta' \to -\infty$, a fact which will be relevant in what follows. 

\begin{figure}[!h]
    \centering
    \includegraphics[width=0.6\textwidth]{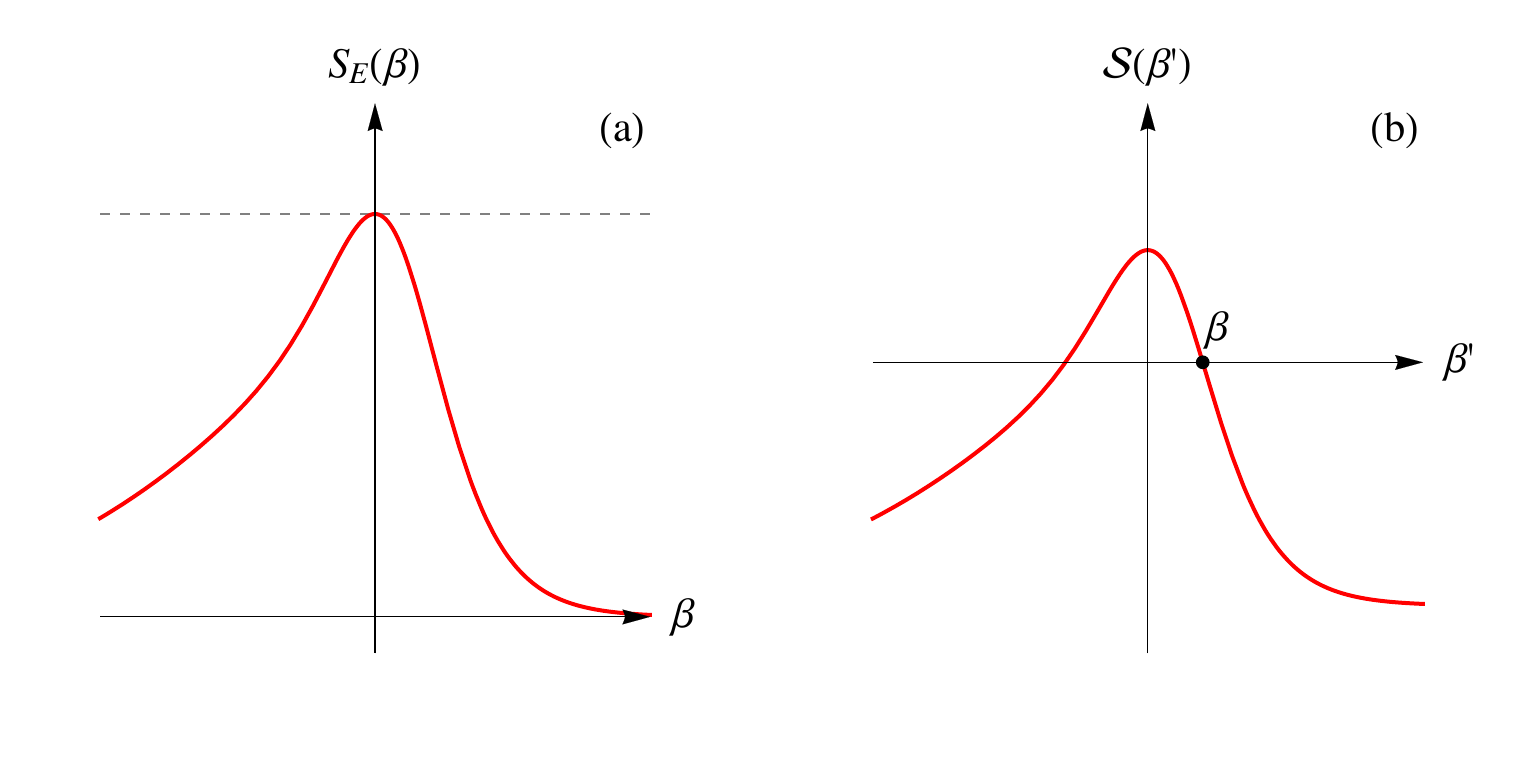}
    \caption{\label{fig:SM_entropies}Similar to Fig.~\ref{fig:SM_energies}, but for the entropies of the environment}
\end{figure}

Next we carry out a similar analysis for the entropy of the environment. 
The situation is depicted in Fig.~\ref{fig:SM_entropies} where we see that $S_E(\beta)$ is  monotonically decreasing for $\beta>0$, but increasing for $\beta < 0$ (Fig.~\ref{fig:SM_entropies}(a)). 
Its maximum therefore occurs at $\beta = 0$ (corresponding to $T = \infty$). 
The function $\mathcal{S}(\beta') = S_E(\beta') - S_E(\beta)$, on the other hand, is again simply a shifted version of the same function (Fig.~\ref{fig:SM_entropies}(b)), which also  crosses the horizontal axis  at $\beta' = \beta$.

\begin{figure}[!h]
    \centering
    \includegraphics[width=0.5\textwidth]{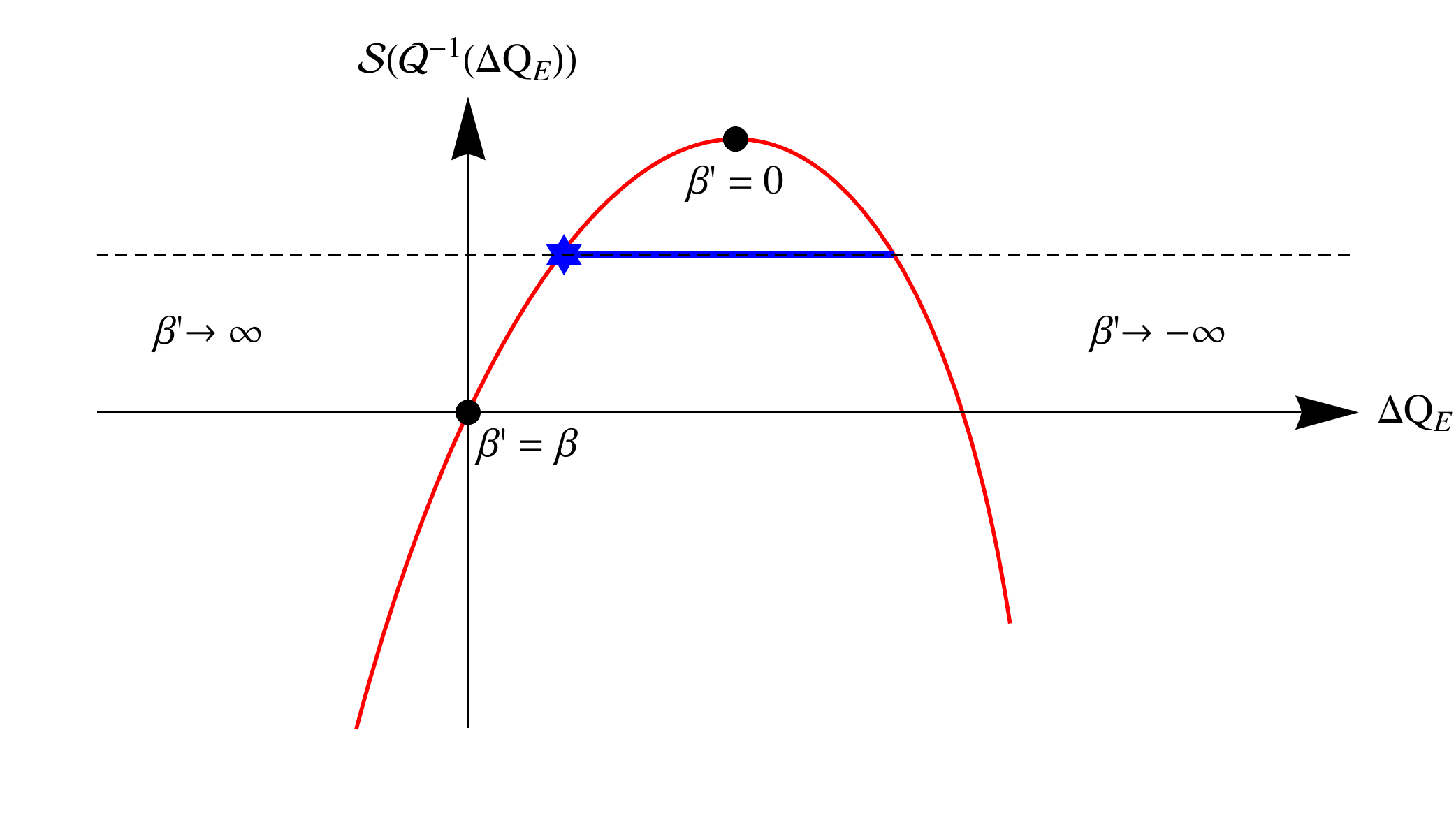}
    \caption{\label{fig:SM_logic}
    The function $\mathcal{S}(\mathcal{Q}^{-1}(\Delta Q_E)$ as a function of $\Delta Q_E$. $\Delta Q_E$ must be in the blue interval because of Eq.~(\ref{landauer_new_0})
    }
\end{figure}

Armed with this intuition, we can now consider Eq.~(\ref{landauer_new_0}) of the main text:
\begin{equation}
    \mathcal{S}(\mathcal{Q}^{-1}(\Delta Q_E)) \geq - \Delta S_S. 
\end{equation}
A sketch of the left-hand side, as a function of $\Delta Q_E$, is shown in Fig.~\ref{fig:SM_logic}.
As seen in Fig.~\ref{fig:SM_energies}(c), lower $\Delta Q_E$ implies higher values of $\beta$. 
Now consider a given value of $-\Delta S_S$, which we denote by a dashed horizontal line in Fig.~\ref{fig:SM_logic}. 
In this case there will be two solutions for $\mathcal{S} =  - \Delta S_S$, corresponding to the two possible inverses of $\mathcal{S}$. 
However, we see that in order to obtain a \emph{lower} bound to $\Delta Q_E$, we need to consider the solution which is to the leftmost part of the plot. 
Since the left part refers to larger $\beta'$, we then conclude that the inverse we should pick is that with $\beta'>0$. 

\section*{Derivation of Eq.~(\ref{gapped})}

We consider Eq.~(\ref{heat_capacity}) and assume $C_E = b e^{-\delta/T}$. 
Moreover, we focus on $T = 0$. 
We may then write
\begin{IEEEeqnarray}{rCl}
    \mathcal{Q}(T') &=& b \delta \;  \bigg[ \frac{T'}{\delta} e^{-\delta/T'} - \Gamma(0,\delta/T')\bigg], \\[0.2cm]
    \mathcal{S}(T') &=& b \; \Gamma(0,\delta/T'),
\end{IEEEeqnarray}
where $\Gamma(a,z)$ is the upper incomplete Gamma function. 
For small temperatures an asymptotic expansion of $\Gamma(0,\delta/T')$ yields 
\begin{IEEEeqnarray}{rCl}
    \mathcal{Q}(T') &\simeq& \frac{b T'^2}{\delta}e^{-\delta/T'}, \\[0.2cm]
    \mathcal{S}(T') &\simeq& \frac{b T'}{\delta} e^{-\delta/T'}.
\end{IEEEeqnarray}
The inverse of $\mathcal{S}(T') = - \Delta S_S$ may, in this case, be written as 
\begin{equation}
    T' = \frac{\delta}{W_0 (-b/\Delta S_S)}, 
\end{equation}
where $W_0(x)$ is the principal solution of Lambert's Product-Log function. 
Eq.~(\ref{landauer_new}) then becomes
\begin{equation}
    \Delta Q_E \geq \delta \frac{(-\Delta S_S)}{W_0(-b/\Delta S_S)}.
\end{equation}
When $b \gg -\Delta S_S$ we can use the asymptotic expansion $W_0(x) \sim \ln x$, which then finally leads to Eq.~(\ref{gapped}). 

\end{document}